\documentclass[sigconf]{acmart}
\usepackage{booktabs}
\usepackage{array}
\usepackage{multirow}

\copyrightyear{2026}
\acmYear{2026}
\setcopyright{cc}
\setcctype{by}
\acmConference[IDC '26]{Proceedings of the 25th Interaction Design and Children Conference}{June 22--25, 2026}{Brighton, United Kingdom}
\acmBooktitle{Proceedings of the 25th Interaction Design and Children Conference (IDC '26), June 22--25, 2026, Brighton, United Kingdom}
\acmDOI{10.1145/3773077.3813777}
\acmISBN{979-8-4007-2283-7/2026/06}

\begin{document}

\title{Mapping the Design Space for Youth Social Media:\\A Framework Centered on Friendship Building}

\author{JaeWon Kim}
\affiliation{%
  \institution{University of Washington}
  \department{The Information School}
  \city{Seattle}
  \state{Washington}
  \country{USA}
}
\email{jaewonk@uw.edu}

\begin{abstract}
This dissertation develops a design framework for friendship-supportive youth social media. I conducted a qualitative meta-analysis across my formative, case-study, and co-design work with teens and young adults, synthesizing recurring design themes into three pillars: \textbf{\textit{social understanding}} (legible norms, intentions, trust, reciprocity, and accountability), \textbf{\textit{placeness}} (spatial and embodied affordances that make online interaction feel inhabitable), and \textbf{\textit{identity alignment}} (authentic expression that remains current, plural, and interpretable). The framework is grounded in interpersonal, developmental, and sociotechnical theory, but its contribution is design-oriented: it translates broader accounts of friendship and social development into the specific ways social media platforms can shape youth friendship building. I initially validate parts of this framework through WhoamI Today (WIT), a platform deployed with 99 youth across the United States and Korea. My proposed work extends this validation through a follow-up deployment while refining the framework as a roadmap for cumulative design research on youth social media.
\end{abstract}



\maketitle

\section{Introduction}
Teenagers worldwide report unprecedented loneliness~\cite{officeofthesurgeongeneral(osg)2023OurEpidemic}, and the role of social media in this epidemic remains deeply contested. Social media is now a central setting for adolescent life, where young people maintain friendships, explore identities, and build relationships across distance~\cite{davis2013}. Yet platforms organized around image-driven feeds, quantified engagement, and influencer culture often intensify social comparison while offering limited support for the reciprocity peer relationships require~\cite{davis2012}.

A common response has been to restrict youth access or rely on individual-level interventions such as screen-time limits and privacy controls. Preventing harm, however, is not the same as creating conditions for flourishing~\cite{kim2024positech, kim2025positech}. Teens themselves report wanting not less time online but more meaningful time in spaces that encourage authenticity, connection, and deeper engagement~\cite{davis2025, landesman2024, alluhidan2024teen, kowalkowski2025belonging}. Even youth who have not experienced direct harm describe ambient anxiety that limits their willingness to connect---a gap the COVID-19 pandemic made especially visible when digital spaces became essential for sustaining friendships.

My dissertation takes this tension as a design problem. Rather than asking only how social media might be made less harmful, I ask what social media would need to provide to actively support friendship building. I use \textit{friendship-supportive design} as a working label for design that creates conditions under which youth friendships can form, deepen, and be sustained; the substance of what those conditions are is what the framework developed in this paper sets out to specify. This framing leads to the central question of my dissertation: \textit{What design principles, organized across what dimensions, constitute a design space for youth social media that actively supports friendship building?}

\section{Framework Development}
I conducted a qualitative meta-analysis across six studies spanning case studies, co-design, surveys, diary studies, interviews, and a field deployment with youth ages 13--25. The first five studies generated the initial framework; the WIT deployment is included as an initial validation and refinement study because it tested selected branches and surfaced tensions among them. Rather than treating the studies as separate case summaries, I re-analyzed them as a corpus of design evidence about how social media becomes a context for friendship building. I extracted 209 design-relevant insights and analyzed them for recurring themes, tensions, and design principles. Interpersonal, developmental, and sociotechnical theory served as sensitizing lenses; the resulting framework specifies how those relational dynamics appear in social media contexts and what designers might build toward.

The six studies contributed complementary evidence. Study A examined BeReal as a case of platform-designed authenticity, showing how synchronized prompts, ephemerality, reciprocity, and audience curation shift norms toward casual sharing while also revealing the limits of unfiltered photo sharing~\cite{kim2024bereal, li2025bereal}. Study B examined teens' dysfunctional fear around privacy and evaluated ten design prototypes that reframe privacy as a social norm rather than a matter of individual vigilance, surfacing affordances that reduce ambient anxiety around audience reach, hostility, and personal missteps~\cite{kim2025privacy, zhao2022}. Study C examined teens' trust-based boundary regulation, showing the need for designs that reduce communication ambiguity, support trust calibration, and make disclosure socially safer~\cite{kim2025trust, akter2025calculating}. Study D was a Discord case study motivated by earlier participants repeatedly naming Discord, unprompted, as friendship-supportive; it showed how a mostly 2D environment can create spatial and third-place experiences through persistent segmented spaces, presence and membership cues, shared activity contexts, social density management, and community-controlled norms~\cite{kim2025discord}. Study E used Fictional Inquiry to move youth outside mainstream platform constraints; participants imagined remote interaction as ambient, organic, low-intensity, embodied, and spatially intelligible rather than feed-based~\cite{kim2026hogwarts, wilson2002six, hall1968proxemics}. Study F deployed WIT to test whether designs derived from selected branches could shift friendship-building behavior in practice and to identify where the framework needed further refinement.

This meta-analysis produced the three-pillar, nine-dimension framework summarized in Figure~\ref{fig:matrix}. The framework is a map of design spaces rather than a checklist of features: a single feature may address multiple dimensions, and an intervention that strengthens one branch may create tensions in another. Naming these branches helps researchers locate the purpose of an intervention, compare platforms at the level of design goals rather than surface features, and identify areas where empirical design knowledge remains thin.

\begin{figure*}[t]

\centering


\begin{tabular}{@{}p{3.8cm}p{4.9cm}cccccc@{\hspace{3pt}}|@{\hspace{3pt}}c@{}}

\toprule

\textbf{Pillar} & \textbf{Sub-Dimension} & \textbf{A} & \textbf{B} & \textbf{C} & \textbf{D} & \textbf{E} & \textbf{F} & \textbf{W1} \\

\midrule

\multirow{3}{3.6cm}{\textit{Social Understanding}\\\footnotesize Interpersonal legibility for navigating interaction}

 & Interaction norms                    & \textbullet & \textbullet & \textopenbullet & & & \textbullet & \textbullet \\

 & Interaction cues and scaffolding     & \textopenbullet & \textbullet & \textbullet & & & \textbullet & \textbullet \\

 & Social accountability and governance & & \textbullet & \textopenbullet & \textbullet & & & \\

\midrule

\multirow{3}{3.6cm}{\textit{Placeness}\\\footnotesize Spatial and embodied affordances for inhabitable interaction}

 & Third place and community            & & & & \textbullet & \textbullet & & \textopenbullet \\

 & Boundaries and personal spaces       & & \textbullet & \textbullet & \textopenbullet & \textbullet & \textbullet & \textbullet \\

 & Shared presence                      & & & & \textopenbullet & \textbullet & & \textbullet \\

\midrule

\multirow{3}{3.6cm}{\textit{Identity Alignment}\\\footnotesize Alignment of self, expression, and others' interpretation}

 & Identity currency                    & \textopenbullet & & \textopenbullet & & & \textopenbullet & \textbullet \\

 & Identity plurality                   & \textopenbullet & \textopenbullet & & & \textbullet & \textbullet & \textbullet \\

 & Relational identity signals          & & & \textopenbullet & \textbullet & \textopenbullet & & \textbullet \\

\bottomrule

\end{tabular}

\caption{Mapping of dissertation studies to the nine framework sub-dimensions. \textbullet{} = study contributed substantially; \textopenbullet{} = partial contribution. Studies: \textbf{A}~BeReal case study (N=29 teens)~\cite{kim2024bereal}; \textbf{B}~teen dysfunctional fear and privacy-as-social-norm study (N=19 co-design + N=136 design evaluation survey)~\cite{kim2025privacy}; \textbf{C}~trust-enabled privacy study (N=19 entry interviews + 7-day diary + co-design exit interviews)~\cite{kim2025trust}; \textbf{D}~Discord case study (N=25)~\cite{kim2025discord}; \textbf{E}~Fictional Inquiry co-design (N=23)~\cite{kim2026hogwarts}; \textbf{F}~WIT crossover field deployment (N=99 across U.S. and South Korea). Proposed: \textbf{W1}~follow-up WIT deployment.}

\Description{A 3-by-9 matrix showing how six completed dissertation studies and one proposed deployment map onto the nine sub-dimensions of the framework. Filled circles indicate substantial contribution; open circles indicate partial contribution.}

\label{fig:matrix}

\end{figure*}

\section{A Three-Pillar Framework for Friendship-Supportive Design}

\textbf{Social understanding} concerns the affordances through which platforms make the social situation legible: what behavior is normal, what an action is likely to mean, when initiation or reciprocation feels welcome, how trust can be calibrated, and how users or communities can respond when expectations are violated. This pillar encompasses interaction norms, interaction cues and scaffolding, and social accountability and governance. Studies A and F showed that posting templates, synchronized prompts, and default interaction structures shape what feels like normal participation, while Studies B and C showed how ambiguous norms, reactions, and audiences erode the trust necessary for self-disclosure, and how dysfunctional fear can paralyze teens even in the absence of direct harm~\cite{kim2025trust, kim2025privacy, akter2025calculating, dym2020social}.

\textbf{Placeness} concerns the spatial, embodied, and environmental affordances that make a platform something youth can inhabit rather than merely a stream of content. Third-place experience is one branch, not the pillar as a whole. More broadly, placeness asks how design lets users understand where they are, who else is present, how public or private an interaction feels, and what forms of movement, boundary, ambience, and co-presence are possible~\cite{meyrowitz1986no, leonardi2015ambient}. Study D showed that Discord can produce spatial experience in a mostly 2D environment through persistent channels, presence and role cues, shared activities, and community-controlled norms~\cite{kim2025discord, fiesler2018reddit}. Study E extended this pillar by showing that youth imagined embodied navigation, ambient co-presence, personal spaces, shared rituals, and spatially intuitive boundaries~\cite{kim2026hogwarts, wilson2002six, hall1968proxemics}.

\textbf{Identity alignment} concerns the fit between an evolving self, the expressive surfaces a platform provides, and the social interpretations those expressions invite. It encompasses identity currency, identity plurality, and relational identity signals. Relational identity signals are pieces of identity and social context---shared interests, roles, mutual affiliations, current activities, and visible traces of participation---that help youth display what matters to them authentically and interpret one another accurately~\cite{walther2002cues}. These signals matter because youth often bond through shared interests and perceived similarity, but they support friendship only when they feel authentic to the sharer and legible to the viewer~\cite{pyle2023social}. Across studies, youth wanted self-presentation that felt accurate rather than performative~\cite{kim2024bereal}, intentional rather than algorithmically inferred~\cite{kim2025trust}, and socially interpretable through shared context rather than aesthetic signaling alone~\cite{kim2026hogwarts}.

\section{Initial Validation Through Deployment}
\label{sec:deployment}

To move from framework development to empirical validation, I built WhoamI Today (WIT)~\footnote{\href{https://linktr.ee/jaewonk}{https://linktr.ee/jaewonk}}, a social media platform that instantiates principles from selected framework branches. WIT was deployed with 99 youth ages 15--25 across the United States and Korea in a four-week crossover study. Participants alternated between an experimental version with friendship-supportive features---daily prompts, targeted prompt-sending, emoji reactions without quantification, private-only comments, a profile-list feed, and close-friends default privacy---and a control version resembling Instagram.

The deployment was not intended to validate the entire framework at once. Rather, it tested whether designs tied to particular branches could shift friendship-building behaviors while revealing where the framework needed refinement. Daily prompts reframed posting as communal participation rather than individual performance, supporting \textit{interaction norms} and \textit{identity currency}~\cite{popowski2024commit}. Directed prompt-sending communicated intent and care while providing social cover for reaching out, supporting \textit{interaction cues and scaffolding}. The profile-list feed gave each user a contained profile-space rather than pushing every post into others' feeds, reducing anxiety around ``taking up space'' and supporting \textit{boundaries and personal spaces}~\cite{zhang2025burst}. Across analyses, participants on the experimental version posted roughly twice as often as on the control despite receiving fewer reactions on average, suggesting that the design supported intrinsically motivated sharing rather than engagement-driven posting.

The same deployment clarified gaps and tensions. Question-sending worked well for existing friends but often felt too direct for weak ties. Private comments reduced performance pressure but also hid social activity that can help a space feel inhabited. The profile-list feed supported self-contained disclosure but reduced passive exposure to peripheral ties, which participants valued for casual familiarity and serendipitous conversation. These findings reinforced the value of treating the framework as an evolving map: deployments can confirm the relevance of particular branches, test designs within them, and reveal collisions or absences across branches.

\section{Proposed Work and Expected Contributions}

My remaining work continues to refine the framework while validating additional branches through deployment. The follow-up WIT deployment (W1 in Figure~\ref{fig:matrix}) tests two claims surfaced by gaps in the first deployment. The first is that new connection formation can be supported without collapsing intimacy gradients, through multi-tiered relationship categories, profile-embedded identity signals that surface shared context, and lightweight followship mechanisms that lower the stakes of initial contact. These features primarily test \textit{relational identity signals}, \textit{interaction cues and scaffolding}, and \textit{boundaries and personal spaces}. The second is that digital environments can better support placeness through ambient, peripheral design~\cite{leonardi2015ambient}. To test this, I introduce a persistent home-screen widget that scaffolds shared social rituals and ongoing awareness of others between active interactions.

The final synthesis will consolidate all completed and proposed work into the framework previewed here. For each pillar and sub-dimension, I will specify relevant theoretical constructs, empirically grounded design principles, example features, and possible measures. The goal is not to claim that one platform can solve youth social media, but to make the design space legible enough that future studies can build cumulatively, compare interventions against shared goals, and identify where evidence remains thin.

One limitation of my deployment-based validation is that it pays less attention to the \textit{social accountability and governance} branch than to the others. Governance is essential to friendship-supportive social media, and substantial research already exists in adjacent areas~\cite{fiesler2018reddit, dym2020social}. However, because WIT is a bounded research platform, its deployments did not grow to a scale where mechanisms beyond basic content and user reporting could be meaningfully tested. I therefore position this branch as part of the framework and as an area where my work should connect to researchers developing richer models of moderation, accountability, and community governance.

The expected contribution of this dissertation is a theory-informed and empirically grounded framework for friendship-supportive youth social media design. Rather than offering platform-specific recommendations, the framework maps the conditions researchers and designers need to consider when designing for friendship building. In doing so, the dissertation reframes youth social media not as a fixed category of harm or benefit, but as an underdeveloped design space where better evidence and more targeted design attempts are still needed.

\begin{acks}
I would like to acknowledge the CERES Network, University of Washington Global Innovation Funds (GIF), and Student Technology Funds (STF), which provided support for this work. This work was also funded in part by the Paul G. Allen School of Computer Science \& Engineering Endowed Fund for Excellence and a gift from Google. I thank my advisor, Alexis Hiniker, and committee members Katie Davis, Casey Fiesler, Amy X. Zhang, and Sean Munson for their guidance throughout this work.
\end{acks}

\bibliographystyle{ACM-Reference-Format}
\bibliography{references, references1, references2, references3}

\end{document}